\documentclass{article}
\usepackage{graphicx}
\begin{document} 
\title{{\bf Binary tree summation Monte Carlo simulation for Potts models}} 
\author{Jian-Sheng Wang\\
{\small {\tt wangjs@cz3.nus.edu.sg}}\\ 
\\
{\small Singapore-MIT Alliance and Department of Computational Science,}\\ 
{\small National University of Singapore, Singapore 119260, Republic of Singapore}}
 
\date{2 June 2002}

\maketitle            

\begin{abstract}
In this talk, we briefly comment on Sweeny and Gliozzi methods,
cluster Monte Carlo method, and recent transition matrix Monte Carlo
for Potts models.  We mostly concentrate on a new algorithm known
as `binary tree summation'.  Some of the most interesting features of
this method will be highlighted -- such as simulating fractional
number of Potts states, as well as offering the partition function and
thermodynamic quantities as functions of temperature in a single run.
\end{abstract}

\section{Introduction} 
The Potts model \cite{wu} is not only fascinating in relation to phase
transitions but also an excellent model for constructing new
algorithms for Monte Carlo simulation.  Before the advent of cluster
algorithms, Sweeny \cite{sweeny} in 1983 already had an algorithm that
has much reduced critical slowing down.  Sweeny algorithm is a
heat-bath algorithm in the Fortuin-Kasteleyn representation
\cite{Kasteleyn} of the Potts model as a percolation problem for any
positive real value of the number of Potts state $q$.  Any
thermodynamic averages in the Potts model representation can be
translated into a percolation bond representation \cite{hu}.  Sweeny
algorithm can be efficiently implemented only in two dimensions. In
three and higher dimensions, it becomes too expensive to simulate,
because each move requires a step to identify if two sites belong to
the same or different clusters.  Nevertheless, it is perhaps the only
algorithm that works for any real values of $q$.

Very recently, Gliozzi \cite{gliozzi} introduced a set of different
transition probabilities than that of Sweeny's.  In Gliozzi's rate,
those links that are occupied by bonds or pair of unoccupied sites
that connect the same cluster are resampled with probability $p$ for
occupation, and $1-p$ for empty.  For unoccupied sites that belong to
two different clusters, the probability for occupation by a bond is
$p/q$ and empty $1-p/q$.  The resampling of a large fraction of the
links appears to decorrelate the configuration efficiently.  Gliozzi
claims \cite{gliozzi} that his algorithm does not suffer from critical
slowing down.  We have shown by explicit calculation of the
correlation times that this is not true \cite{wang-oner-swendsen-tau}.
In two dimensions, we have good logarithmic divergence of the
correlation time with system linear sizes for the Ising model.  For
two-dimensional three-state Potts model, we found dynamical critical
exponent $z \approx 0.5$, and for the three-dimensional Ising model,
the exponent is about 0.4.

The Swendsen-Wang \cite{Swendsen-Wang} and Wolff \cite{Wolff}
algorithms work only for integer values of $q$.  They can be
implemented very efficiently.  For the Swendsen-Wang algorithm, each
sweep takes CPU time of $O(N)$, and for Wolff algorithm, it is
proportional to the size of the cluster that is being flipped.  They
reduce but not completely eliminate critical slowing down at the
critical temperature.

The traditional methods mentioned above simulate at one fixed
temperature $T$, thus the results are in terms of discrete points.  A
curve is obtained by connecting these points.  The transition matrix
Monte Carlo \cite{Wang-Swendsen-JSP} and other related algorithms
\cite{ferrenberg,Berg,oliveira,wang-landau} generate continuous function of
temperature or other variable from a single simulation, thus greatly
enhanced efficiency.  The multi-histogram sampling of Weigel et al
\cite{weigel} in the context of $q$-state Potts model should be
mentioned here.  The multicanonical simulation method of Berg
\cite{Berg,berg-review} also has the additional advantage of much
reduced correlation times for first-order phase transitions.

It is very attractive to have algorithms that do not have correlation
at all.  In fact, there are such algorithms, such as that of random
percolation.  Since the samples are drawn at random for each
configuration, each configuration is independent of the previous
configurations.  The simple sampling method of Hu \cite{hu-potts} has
this property.  Newman and Ziff \cite{Newman-Ziff} gave an efficient
implement that work by exact weighting of the probability $p$ so that
after the simulation, the whole curve $Q$ of some physical observable
versus $p$ can be computed.  In this paper, we present some attempts
\cite{wang-oner-swendsen-Goergia} that generalize Newman-Ziff method
to general Potts model of $q$ states.  These algorithms all have the
common feature that bonds are added to the system one by one.  Once
there are on the lattice, they do not move.  It is a nonequilibrium
process in the sense that there is no (Monte Carlo) time translation
invariance for the process.  The rest of the paper is organized as
follows: in the next section, we briefly review the Newman-Ziff
method, followed by generalization of the partition function to Potts
model.  We then discuss a number of algorithms, and show their
efficiencies and also point out their shortcomings.  We conclude in
the last section.
 
\section{Newman-Ziff-type methods}
Consider the simple bond percolation problem.  We generate
configurations by putting bonds at every nearest neighbor link of a
$d$-dimensional hypercubic lattice with probability $p$, and absence
of bond with probability $1-p$.  Any average of physical quantity
$Q(\Gamma)$ of configuration $\Gamma$ can be computed exactly from
\begin{equation}
 \langle Q \rangle = \sum_{\Gamma} Q(\Gamma) p^b(1-p)^{M-b},
\label{eq1}
\end{equation}
where the summation is over all the $2^{M}$ configurations; $b$ is
number of bonds present, and $M=Nd$ is maximum possible number of
bonds.

In a standard simulation, one fixes a value of $p$, and generates
configurations by visiting each link and putting a bond with
probability $p$.  Estimates of $\langle Q \rangle$ are obtained by
taking sample means of quantity $Q(\Gamma)$.  Newman and Ziff
\cite{Newman-Ziff} instead considered a computation of $Q_b$ for each value of
$b$, and then reconstructed from $Q_b$ the function $\langle Q
\rangle$ of $p$.  From Eq.~(\ref{eq1}), we have
\begin{equation}
   \langle Q \rangle = \sum_{b=0}^{M} \sum_{\Gamma_b} Q(\Gamma_b) 
    p^b(1-p)^{M-b} = \sum_{b=0}^{M} p^b(1-p)^{M-b} 
    { M! \over b! (M-b)! } Q_b,
    \label{eq2}
\end{equation}
where $Q_b$ is defined as average over all the $M!/(b! (M-b)!)$
configurations $\Gamma_b$ of exactly $b$ number of bonds,
\begin{equation}
   Q_b = { b! (M - b)! \over M! } \sum_{\Gamma_b} Q(\Gamma_b).
\end{equation}

Newman and Ziff proposed the following algorithm to evaluate $Q_b$.
Each sweep of the lattice starts with empty lattice, putting bonds one
by one at random over these that is empty.  This will generate $M+1$
configurations with 0, 1, $\cdots$, $b$, $\cdots$, $M$ number of
bonds.  From them, we calculate $Q_b$ for each $b$.  With the help of
Hoshen-Kopelman algorithm \cite{hoshen}, the whole calculation of one
sweep can be done in CPU time of $O(N)$.  Although configurations
within a sweep are highly correlated, each sweep is an independent
one.  Very accurate percolation threshold was obtained this way.

To generalize this procedure to Potts model, we note that the
analogous equation to Eq.~(\ref{eq2}) is
\begin{equation}
   \langle Q \rangle = \sum_{b=0}^{M} \sum_{\Gamma_b} Q(\Gamma_b) 
    p^b(1-p)^{M-b} q^{N_c(\Gamma_b)}= 
    \sum_{b=0}^{M} p^b(1-p)^{M-b} 
    c_b\, Q_b,
\end{equation}
where $N_c(\Gamma)$ is the number of clusters of the configuration
$\Gamma$, and
\begin{equation}
c_b = \sum_{\Gamma_b} q^{N_c(\Gamma_b)}, \qquad
Q_b = {1\over c_b} \sum_{\Gamma_b} Q(\Gamma_b)\, q^{N_c(\Gamma_b)}.
\end{equation}
The probability $p$ is related to temperature $T$ by $p = 1 -
\exp\bigl(-J/(kT)\bigr)$.  When $q \neq 1$, $c_b$ is no longer known
exactly, we must compute by Monte Carlo simulation; the quantity $Q_b$
is now an average over the configurations of a given number of bonds
$b$ distributed not uniformly but according to $q^{N_c(\Gamma_b)}$.
This makes the simulation much harder.

\section{Algorithms for computing $c_b$ and $Q_b$}
\subsection{simple surviving and dying process}
The following algorithm, although very inefficient, generates correct
probability distribution for the samples, and introduces the way $c_b$
and in general any $Q_b$ can be computed.

Starting with an empty lattice, one sweep consists of repeated
application of the following steps until the process dies:
 
\begin{enumerate} 
 
\item Pick an unoccupied neighbor pair at random for the next bond. 
 
\item If inserting a bond   
 \begin{enumerate} 
\item does not change the cluster number ($\Delta N_c=0$), accept the   
       configuration; 
\item merge two clusters, so that the cluster number decreases by 1,  
       ($\Delta N_c = -1$), accept the configuration with probability $1/q$, 
        or reject the configuration and terminate the process 
       (and begin the next sweep from an empty lattice). 
  \end{enumerate} 
 
\item Take statistics of the survival configurations (with equal weights). 
\end{enumerate} 
The probability distribution of the process at $b$ number of bonds is
proportional to $q^{N_c(\Gamma)}$.  The configurations that have equal
number of clusters from $b$ to $b+1$ bonds have the same probability,
but these that merge clusters appear with a probability smaller by a
factor $1/q$.  The cumulative effect gives the desired probability
distribution.  Unfortunately, since the number of samples is
exponentially small for large $b$, the algorithm is of only conceptual
use.

The number $c_b$ can be related to sample average of the conditional
survival probability.  Let $n_0$ be the number of empty links that
connect same cluster, and $n_1$ be the number of empty links that are
on different clusters.  Then the conditional survival probability is
\cite{wang-oner-swendsen-Goergia}
\begin{equation}
\lambda_b = \left\langle{  n_0 + n_1/q \over n_0 + n_1 }\right\rangle 
           = {(b+1)\, c_{b+1} \over (M-b)\, c_b}.
\end{equation}
  
\subsection{N-fold way}
To speed up the simulation, we prevent the process from dying, but the
price we have to pay is that we need to give weight to samples.
Specifically, we follow the method of $N$-fold way: we choose a type-0
class with probability $n_0/(n_0+n_1/q)$ or type-1 class with
probability $n_1/q/(n_0 + n_1/q)$.  Once we have decided the class, we
pick an empty link from among $n_0$ bonds (for type-0 class) or among
$n_1$ bonds (for type-1 class) at random.  The probability
distribution of such a process is not exactly what we wanted, but
rather it is
\begin{equation}
  P_{{\rm path}}(\Gamma_b) = { 1 \over \prod_{b'=0}^{b-1} A(\Gamma_{b'}) }
    q^{N_c(\Gamma_b) - N_c(\Gamma_0)},
\end{equation}
where $A(\Gamma) = n_0(\Gamma) + n_1(\Gamma)/q$, and $N_c(\Gamma_0) =
N$ is the number of clusters of the empty lattice.  To compute the
desired average we must weight with $A_b^{{\rm tot}} =
\prod_{b'=0}^{b-1} A(\Gamma_{b'})$,
\begin{equation}
  Q_b = {\displaystyle \sum_{\Gamma_b} q^{N_c(\Gamma_b)} Q(\Gamma_b)  \over
         \displaystyle \sum_{\Gamma_b} q^{N_c(\Gamma_b)} } \approx 
      {\displaystyle \sum_{{\rm MC\ samples}} A_b^{{\rm tot}} Q(\Gamma_b) \over
       \displaystyle \sum_{{\rm MC\ samples}} A_b^{{\rm tot}} } .
\end{equation}
Due to fluctuations in the weights, inefficiencies are unavoidable.
In fact, the above method can give realizable results only for small
sizes of order 10.

\subsection{binary tree summation}
To overcome the above problem, we propose the following method
\cite{wang-oner-swendsen-Goergia} named binary tree summation Monte
Carlo.  For each sweep, the algorithm consists of two parts: the
simulation part and summation part.  In the simulation part, we always
pick a type-1 link (unoccupied link that bridges two clusters) at
random, thus always merge clusters in each step.  The probability
distribution of the configuration sequence generated is
\begin{equation}
 P(\Gamma_0 \to \Gamma_1 \to \cdots \to \Gamma_i) = 
   { 1 \over n_1(\Gamma_0) } 
   { 1 \over n_1(\Gamma_1) } 
   \cdots
   { 1 \over n_1(\Gamma_{i-1}) }. 
\end{equation}
In the summation part, we consider all possible ways of inserting
type-0 bonds in between each merge step of simulated configurations.
The configurations are weighted in such a way to realize the desired
probability distribution, $P(\Gamma) \propto q^{N_c(\Gamma)}$.
Specifically, the weight of a configuration which is specified by
number of bonds $b$ and sequence number in the simulation $i$, is the
product of factor $n_1/q$ (for each merge step) and $n_0$ (for each
insertion of type-0 bond that does not merge clusters).  This weight
$w(b,i)$ can be computed recursively for all $b$ and $i$ with computer
time of $O(N^2)$, by
\begin{equation} 
  w(b+1,i) = w(b,i) n_0(b,i) + w(b,i-1) n_1(b,i-1)/q, 
\end{equation} 
where $n_1(b,i) = n_1(i)$, $n_0(b,i) = n_0(i) - b + i$, with the
starting condition $w(0,i) = \delta_{i,0}$ and the constraint
$n_0(b,i) \ge 0$.  The value $w(b,i)$ is nonzero only for $b \ge i$.
The computation of the weights $w(b,i)$ is similar to a recursive
computation of the binomial coefficients.  The final statistics of 
a quantity $Q$ can be computed as
\begin{equation} 
  Q_b = \langle W_b \rangle^{-1} 
             \Bigl\langle \sum_{i=0}^{N-1} w(b,i) Q(i) \Bigr\rangle,
\end{equation}  
where $W_b = \sum_{i=0}^{N-1} w(b,i)$, $Q(i)$ is the quantity at the
$i$-th cluster merge step.  The ratio $(b+1)c_{b+1}/((M-b)c_b)$ can be
computed from the expectation value of $(n_0 + n_1/q)/(M-b)$.
 
Note that the number of states $q$ enters into the simulate only
through the weights and is not used in sampling, we can use any real
or complex value for $q$.  Moreover, if we collect appropriate
histogram (of $O(N^2)$ entries), we can reconstruct the equilibrium
average for any $p$ and $q$ through a single simulation.  The details
of the algorithm and some applications will be discussed elsewhere
\cite{wang-kozan-swendsen-BTS-prepare}.

\begin{figure}[t]
\includegraphics[width=0.85\columnwidth]{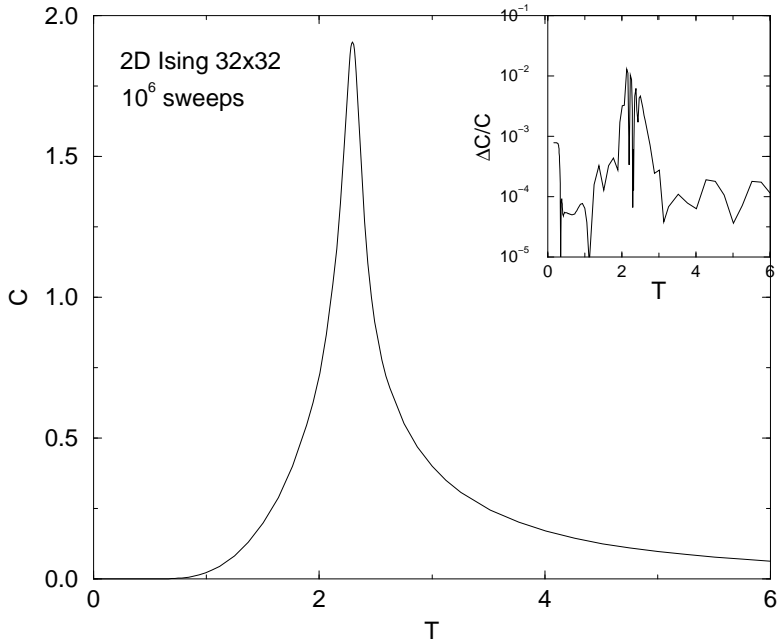}
\caption[C vs T]{Specific heat for the two-dimensional Ising model
computed by binary tree summation method.  The insert gives the
relative error with respect to exact result.  The system size is $32
\times 32$ with $10^6$ Monte Carlo sweeps.}
\label{fig-1}
\end{figure}

\begin{figure}[t]
\includegraphics[width=0.85\columnwidth]{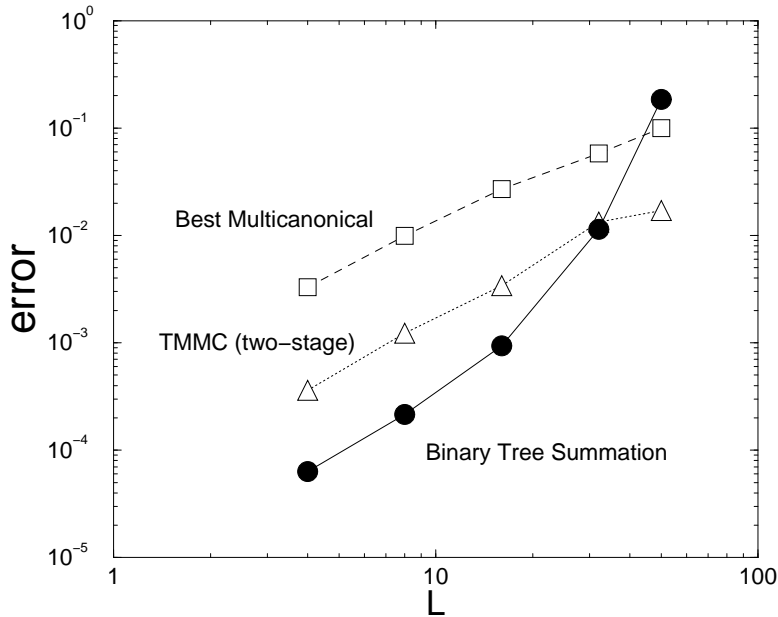}
\caption[epsilon]{Average relative errors in $c_b$ after $10^6$ sweeps
for the two-dimensional Ising model computed by binary tree summation
method (circles), and errors in density of state $n(E)$ from
multicanonical method (squares) and a two-stage transition matrix
Monte Carlo method (triangles).  The data are from
ref.~\cite{wang-oner-swendsen-Goergia} and \cite{Wang-Swendsen-JSP}.}
\label{fig-2}
\end{figure}

In Fig.~\ref{fig-1}, we present a plot of specific heat of the Ising
model on a $32\times 32$ lattice.  Errors in comparison with exact
results \cite{beale} are indicated in the insert.  
For this lattice size, the errors
are quite small for all values of $T$.  In Fig.~\ref{fig-2}, we plot
the average relative errors in $c_b$, defined by
\begin{equation}
   \epsilon = {1 \over M} \sum_{b=0}^M \left|  
                            {c_b \over c_b^{\rm{exact}}} - 1\right|. 
\end{equation}
The quantity $c_b$ plays the role of density of states as considered
in refs.~\cite{kawashima} and \cite{yamaguchi}.  For the binary tree
summation data, we have rescaled the error by a factor $\sqrt{t_{\rm
cpu}/1.9}$ so that the comparison of errors is on an equal footing of
given amount of CPU time.  As we can see from the plot, the errors for
small lattices are much smaller than that of both the multi-canonical
simulation and transition matrix Monte Carlo method
\cite{Wang-Swendsen-JSP}.  However, as the system size becomes larger
than $32 \times 32$, the error becomes comparable and less favorable
to the other methods.

The decreased performance of binary tree summation method for large
systems is partly due to the $O(N^2)$ nature of the algorithm, and
perhaps more importantly, it is not an importance sampling method.
The fluctuation of the total weights $W_b$ in binary tree summation
method is much smaller comparing to the $N$-fold way.  Nevertheless,
the weights introduce extra errors and become important for large
systems.  The $q=1$ percolation problem do not have such problem as in
that case the total weight for each $b$ is a constant.  A ``fudge
factor'' may be introduced in the probability of choosing a type-1
bond with the aim to reduce the total weight fluctuation.  We are
still investigating on such possibilities.

\section{Conclusion}
We first compared the Sweeny and Gliozzi rates for the simulation of
Potts models.  The Sweeny and Gliozzi rates give the same dynamics and
also do not completely eliminate critical slowing down.  We then
introduced Newman and Ziff method for simulating percolation problem.
Our central goal is to simulate Potts models in the Fortuin-Kasteleyn
representation in the spirit of Newman and Ziff.  
Binary tree summation method is our best attempt in this direction.
Although the samples
are generated without correlation between sweeps, we lost the property
of importance sampling.  In this sense, we have not completely solved
the problem, thus the methods are not efficient for large systems.
We hope the present work will stimulate research for algorithms that
have zero correlation, yet realize importance sampling.
 
\section*{Acknowledgements}
The author thanks Prof. Chin-Kun Hu for the invitation for the
presentation of this work.  He acknowledges for collaborations with
Oner Kozan and Robert H. Swendsen and many discussions during a
sabbatical leave at Carnegie Mellon University, where most of the work
discussed in the paper were done.  He also thanks Yutaka Okabe and
Naoki Kawashima for discussions and for hosting a part of a sabbatical
leave at the Tokyo Metropolitan University.

\end{document}